# HYBRID WORKFLOW POLICY MANAGEMENT FOR HEART DISEASE IDENTIFICATION

DONG-HYUN KIM[*1], WOO-RAM JUNG [1], CHAN-HYUN YOUN[1]

[1]Department of Information and Communications Engineering,

Korea Advanced Institute of Science and Technology, Korea.

**ABSTRACT:**

As science technology grows, medical application is becoming more complex to solve the physiological problems within expected time. Workflow management systems (WMS) in Grid computing are promising solution to solve the sophisticated problem such as genomic analysis, drug discovery, disease identification, etc. Although existing WMS can provide basic management functionality in Grid environment, consideration of user requirements such as performance, reliability and interaction with user is missing. In this paper, we propose hybrid workflow management system for heart disease identification and discuss how to guarantee different user requirements according to user SLA. The proposed system is applied to Physio-Grid e-health platform to identify human heart disease with ECG analysis and Virtual Heart Simulation (VHS) workflow applications.

**KEYWORDS:** Grid Workflow, Heart Disease Identification, Physio-Grid, Resource Management.

## I. INTRODUCTION

To date, workflow management systems (WMS) in Grid computing are becoming widespread for scientific applications to solve sophisticated problem such as genomic analysis, drug discovery, disease identification, etc. As scientific applications become more complex, the management of Grid workflow system has become one of the challenging issues. Efforts to solve the above problems are being made in various research centers, e.g., PEGASUS WMS [1], Triana [2], Gridflow, etc. can provide the basic infrastructure to solve scientific workflow application. Although those systems can provide workflow management functionality in Grid environment, consideration of user requirements such as performance, reliability and interaction with user is missing. Moreover, current Grid WMS only consider about workflow itself without considering interaction with user like conventional enterprise WMS. To satisfy user requirements, it is necessary to provide functionality of express working procedure easily such as data retrieval or parameter input. Kepler [3][4], however, has the powerful functions to compose complex scientific application which has complex flow of data from one analytical step to another. Another limitation in conventional workflow management system is lack of supporting resource management functions. As each task in Grid workflow allocated to specific resource, consideration of resource management issue in workflow is important to satisfy user requirements. Moreover, in science Grid application the requirements of user in different domain should be guaranteed cost effective way. Especially in medical Grid, e.g., doctor, patient, resource provider and service provider have different requirements. In this manner, considering SLA [6] of each user is important function in medical Grid application.

Grid workflow allocated to specific resource, consideration of resource management issue in workflow is important to satisfy user requirements. Moreover, in science Grid application the requirements of user in different domain should be guaranteed cost effective way. Especially in medical Grid, e.g., doctor, patient, resource provider and service provider have different requirements. In this manner, considering SLA [6] of each user is important function in medical Grid application.

To address above problems, we came to the idea of integrating existing workflow management tool – Pegasus and Kepler, to Policy Quorum based Resource Management (PQRM) [5] middleware which is termed as a hybrid workflow management system. Policy management mechanism is used to integrate two different workflow management systems in a flexible way and provide advanced management functions to guarantee user requirements. Our WMS applied to Physio-Grid [9] e-health platform to help medical doctor identify human heart disease with Electrocardiogram (ECG) analysis and Virtual Heart Simulation (VHS) [7] workflow applications

[*] Corresponding author email epick4u@gmail.com

## II. MODEL DESCRIPTION OF HYBRID WORKFLOW POLICY MANAGEMENT

*A. Hybrid Workflow Management System for Heart Disease Identification*

Physio-Grid is a Grid E-Health application for heart disease identification which has various ECG and MCG experimental data, Virtual Heart Simulation (VHS) data and clinical data of patients under highly distributed and heterogeneous environment. We apply Hybrid WMS in Physio-Grid to provide medical diagnosis application service with WMS to enhance performance and guarantee user requirements. The Hybrid WMS is form of integrated model of two different kind of workflow - high level WMS and low level WMS. The high level WMS provide functions contain user interaction, data retrieval from database and simple calculation module. On the other hand, low level WMS will provide computing intensive applications such as ECG analysis and virtual heart simulation and automatically submitted to Grid environment for the performance reason. The well-known two different workflow systems is used in our proposed system – Kepler for high level workflow composition and Pegasus for low level (Grid) workflow scheduling and execution. We design and implement integration modules both in Kepler and PQRM middleware with Pegasus engine and adopt policy management mechanism to guarantee user requirement. Policy management provides flexible integration of two different WMS, which is composed of three modules – policy decision point, policy enforcement point and policy repository. Pre-defined policy is enforced to middleware including Kepler and Pegasus according to user requirement which is described in SLA. Figure 1 describes layered architecture of Physio-Grid with Hybrid WMS. The Physio-Grid can be divided by three layers – Application layer, core middleware layer and resource fabric layer. Physio-Grid web portal in application layer provide easy interface of using each middleware components. Service user can specify SLA via Kepler user interface. Concrete workflow which is designed by Kepler is also monitored via the web page. Moreover, the result of workflow is visualized with several graphs or movie clips. Core middleware layer contain workflow application services and Hybrid WMS middleware. There are currently three workflow application in Physio-Grid system to diagnosis heart disease – ECG Analysis only workflow, ECG Analysis and disease detection workflow, and ECG Analysis with Virtual Heart Simulation. Each workflow is selected by SLA specified by user with reasonable cost. These workflow applications automate diagnosis procedure and provide easy interface to service user. PQRM middleware with Pegasus workflow engine used to execute workflow application in Grid network.

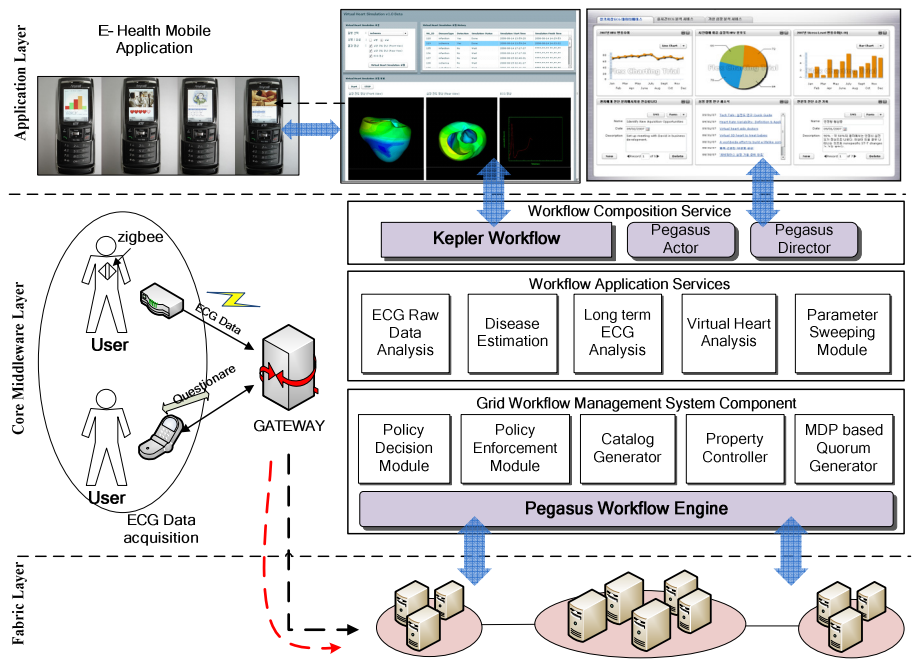

**Figure 1 Architecture of Physio-Grid**

Newly developed three module is used – policy management module, interface of Pegasus low level workflow module, and resource selection module. In high level workflow, its node can call Pegasus workflow engine when computing intensive job or complex job is allocated. Policies selected from policy repository act as cost or performance optimized way to guarantee user requirements.

We have designed heart disease workflow applications in Physio-Grid. Figure 2 show one of the heart disease identification workflow applications which include ECG analysis module, disease estimation, and Virtual Heart Simulation module. Two sub-workflows refer to ECG data Analysis and Virtual Heart Simulation.

These two sub-workflows are designed through Kepler user interface and parsed to DAX format to execute in Pegasus workflow engine which is located in Grid environment. As shown in figure, the output of sub workflow node is passed to high level WMS automatically and simple calculation is performed to analyze the result of the sub workflow node. When sub workflow node is called from Kepler WMS, PQRM resource management system select the candidate resources from Grid network and pass the resources list to Pegasus WMS to run the target application.

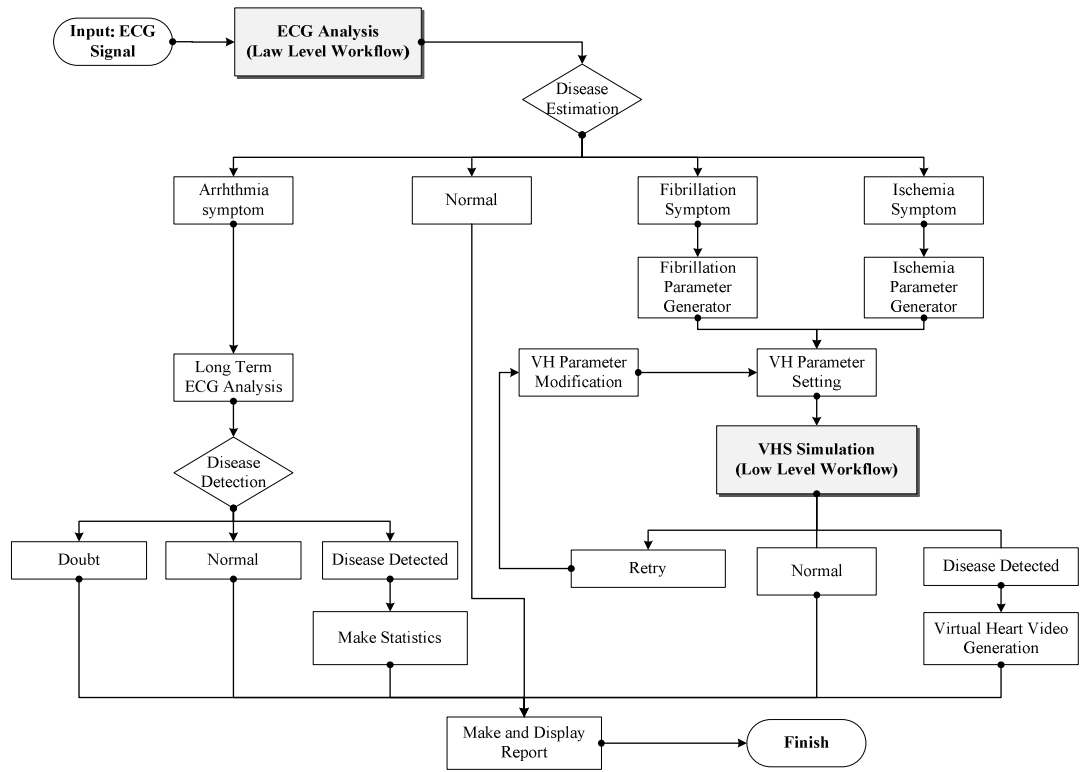

Figure 2 Workflow for Heart Disease identification

The heart disease identification workflow can help doctor to diagnosis human heart disease such as arrhythmia, fibrillation and ischemia. Input data of the workflow is human ECG signal which can be gathered in real-time from patient home environment. After the ECG signal analyzed from ECG analysis sub-workflow, disease estimation procedure is performed according to result symptom. In case of fibrillation or ischemia symptom, advanced diagnosis is required and Virtual Heart Simulation (VHS) which needs huge computing power is executed with pre-defined parameters in Grid environment until the analyzed result match with patient real ECG data. On the other hand, arrhythmia only need long term ECG analysis to diagnosis patient current heart status. These result data will be stored in Physio-Grid integrated database system and medical doctor can see the result from Physio-Grid user portal.

*B. Functional Architecture of Hybrid WMS*

The main design issue of Hybrid WMS is flexible integration of existing WMS. To meet the problem of integration, we divide entire system to several components. Newly implemented module includes PQRM Workflow Management Service (PWMS), Catalog Generator, Property Controller and Pegasus Director, Actor to provide interface with Kepler WMS. In addition, we design and implement Policy Management modules that include Policy Decision Point (PDP), Policy Repository, and Adaptive Policy Enforcement module (PEP). At last, CIM based Information Base provides dynamic information of policy scheme and properties. Figure 3 shows the functional architecture of Hybrid WMS with existing two WMS system and policy integration mechanism. The GLOBUS Toolkit 4.2 [9] is used for core Grid middleware. Above middleware platform Physio-Grid e-health application is implemented to help doctor diagnosis human heart disease.

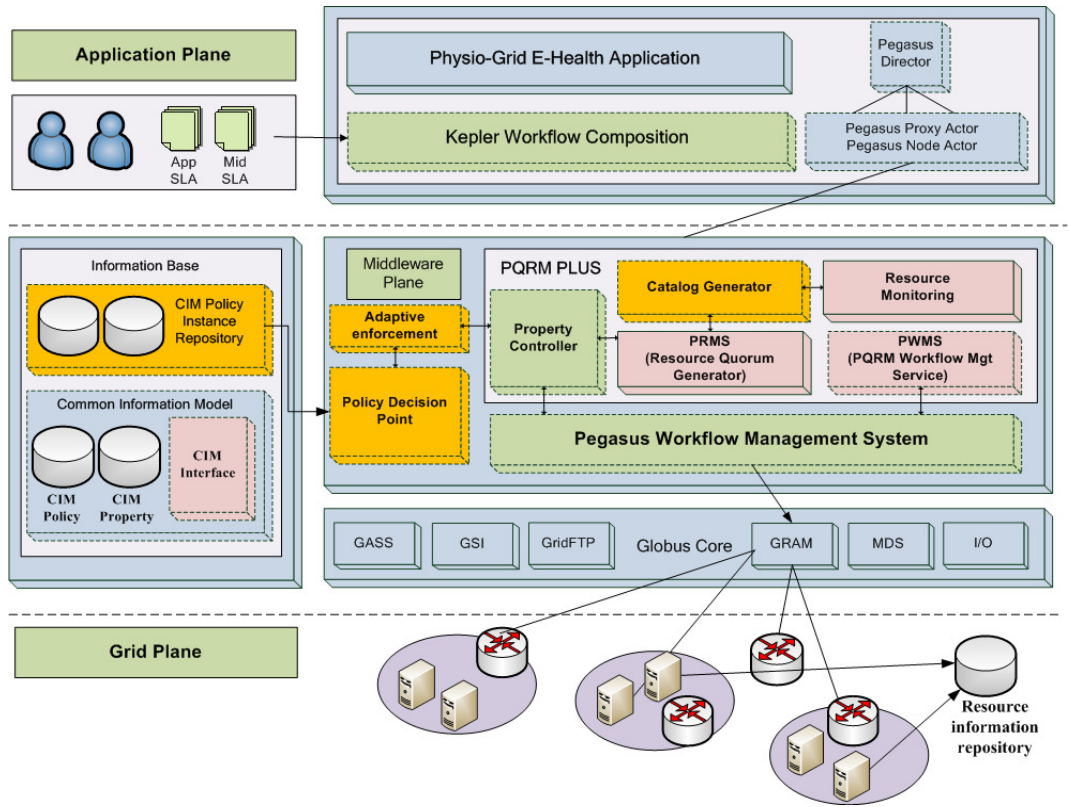

**Figure 3 Functional Architecture of Hybrid WMS**

Followings are the detailed explanation of Hybrid WMS components:

- **Kepler workflow engine**: Kepler workflow engine is used for composite entire workflow with its GUI. We develop our own actor and director to interact with Pegasus and PQRM. Pegasus low level workflow also can composite via Kepler GUI.

- **Pegasus workflow engine**: Pegasus workflow engine is integrated to PQRM middleware to execute computing or data intensive sub workflow in Kepler through Grid environment. Pegasus properties are controlled by policies based on user SLA.

- **Policy Decision Point (PDP)**: Policy is decided on policy decision point. Condition in policy is compared with user SLA and adequate action will be enforced to PQRM middleware. In some specific case, CIM property information will affect enforcement act to increase performance or reliability of the system.

- **Adaptive Policy Enforcement Module**: Adaptive policy enforcement module is actual enforcement point in PQRM. PQRM middleware including Pegasus dynamically change their action behaviors according to action described in chosen policy. The dynamic enforcement of action is implemented by JAVA dynamic class loading technology.

- **PQRM Workflow Management Service (PWMS)**: PWMS is composed of Catalog generator and Property controller. It is Pegasus interface module in PQRM. Catalog is additional information of abstract workflow in Pegasus and PWMS will generate the catalog files automatically. Property controller is involved in configuration of Pegasus and PQRM properties.

- **CIM Information Base:** CIM Information Base provides additional information to policy through CIM policy and CIM property class scheme. Additional information can avoid performance degradation or system fault.

C. *Policy Decision Scheme and Operation Procedure*

Policy decision and its procedure are simple but important in terms of guaranteeing different domain user requirements. We assume that end user of application system do not know detail about service from low level middleware. We can define high level user requirement as soft requirement. Soft requirement can be described as natural language such as "High Performance", "Low Cost", etc. soft requirement is actual constraint of decision making process and tightly related to the preference of the end user. As we consider that our end user might be a medical doctor or patient who is not friendly with computer system, soft requirement is reasonable to describe their high level requirement. We define the definitions for soft requirement as follows.

$$SLA_{app}^k = \{R^k, Pf^k, As^k\} \quad (1)$$

where $R^k$ is required resource set level, which can be a input of the Available Resource Quorum (ARQ) Generator to get the appropriate resource set according to user requirements. Resources in higher level resource set have lower Allocation Cost (AC) than other resource set. AC is derived as follow.

$$AC_{res}^i(t) = \alpha \cdot AC_{net}^i(t) + \beta \cdot AC_{sys}^i(t) \quad (2)$$

Each $AC_{net}^i(t)$ and $AC_{sys}^i(t)$ stands for network status and system status of resource i and $\alpha, \beta$ represent weighting value for each factor. Smaller $AC_{res}^i(t)$ means high performance resources with high cost. $Pf^k$ means required low level workflow performance of user k, which can manipulate scheduling function of low level workflow engine. At last, $As^k$ means required service level of high level application. In other words, it is a service level of different services which can be provided by specific high level workflow applications. To guarantee above user requirement in WMS, we define three kinds of policy - Resource police, low level workflow policy, and high level workflow (application service) policy. The policy set of user k is defined as follows.

$$P_{set}^k = \{P_{as}^k, P_{res}^k, P_w^k\} \quad (2)$$

where $P_{as}^k$ is an application level policy of user k. In the same way $P_{res}^k, P_w^k$ can be described as resource policy and low level workflow policy of user k.

High level WMS refers to Kepler WMS and low level WMS refers to Pegasus WMS. User can interact with workflow management system through Kepler interface to see the chosen policy to decide whether execute the workflow or not. If user satisfies, Kepler workflow will be executed with designed workflow process. During the runtime of workflow process each node of Kepler will check the node type. System will submit task to Pegasus workflow engine if the node type is Grid node, otherwise the task is executed on local machine using Kepler workflow engine. If the last task is detected, the workflow engine will get the output of workflow and terminate. Currently data visualization of the result is not supported in Kepler interface but available on Physio-Grid web interface.

## III. PERFORMANCE EVALUATION

### A. Experimental Result and Discussion

To integrate Hybrid WMS with Physio-Grid System, we define three types of policies– Resource policy, low level workflow policy and application service policy. Among these policies, resource policy and low level workflow policy affect to completion time of entire workflow application.

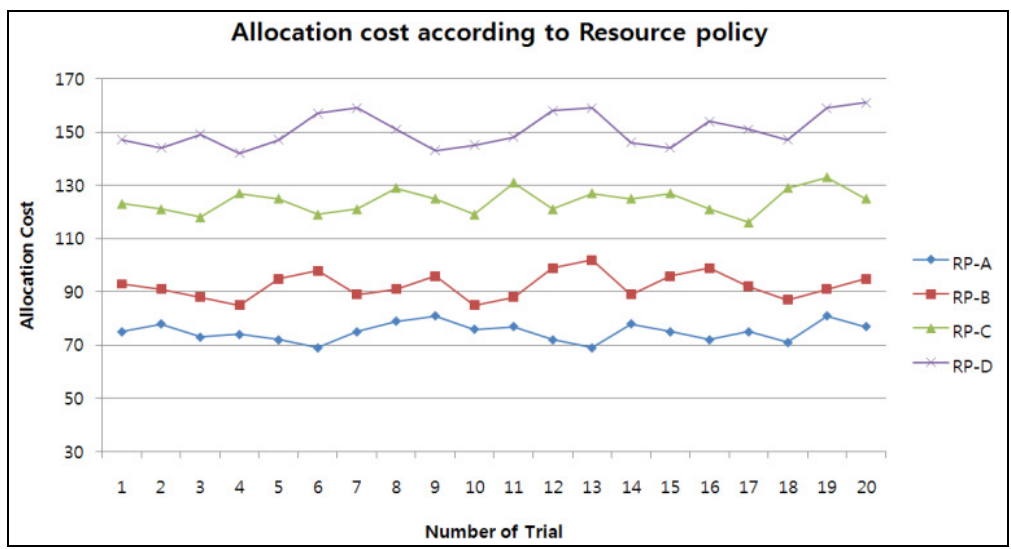

**Figure 4 Average value of allocation cost per hour**

Figure 4 shows average value of allocation cost of six top ranked resources. Owing to dynamic network environment, the cost of each machine changes as time flies. Lower value of allocation cost means high performance resources with high price. In proposed system, PQRM select adequate resource set by the chosen policy. In this figure, RP-A policy chose best resource set.

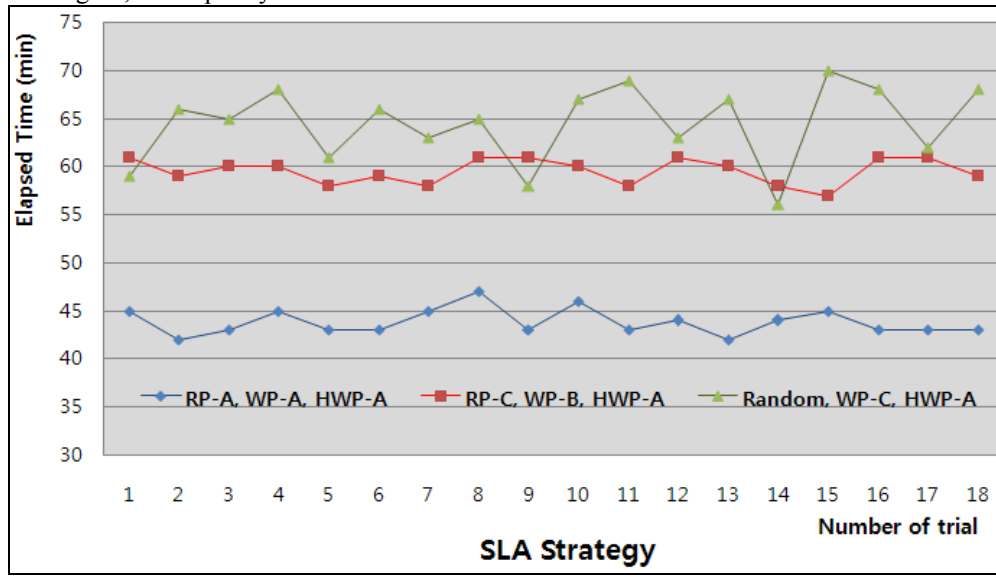

**Figure 5 Affect of different policies**

Figure 5 shows performance effect of the policies. Three different policies is applied before workflow (ECG Analysis and Virtual Heart) is executed. The resource is selected by PQRM resource policy. Same application services are executed simultaneously and submitted to Grid network. As you can see in the figure, the policy set with RP-A(Resource Policy A), WP-A(Workflow Policy A), HWP-A(High Level Workflow Policy A) shows best performance. On the other hand, rest of two experiments shows worse completion time. The 2nd and 3rd policy shows that without resource policy (Randomly selected) the variance of completion time is bigger and unstable.

## IV. CONCLUSION

We propose a Hybrid Workflow Policy Management system to handle SLA guarantee issues in Grid workflow management area. In this proposed system, Grid workflows are managed and scheduled just in-time in terms of a decentralized scheduling architecture which provides a more flexible and loose-coupled control. Furthermore, users' requirements can be guaranteed according to enacted SLAs by utilizing distinct mapping policies which are determined by the policy decision point. We show that integration of existing workflow management system with policy management mechanism can provide advance features compare to conventional workflow management system. This idea combines the policy management, workflow management together and takes advantages from both in order to give better support to Grid computing middleware. Finally, we adopt proposed system to Physio-Grid E-Health System to identify heart disease. In experiment and evaluation, the result shows that proposed system can guarantee user requirement and shows the advanced feature compare with existing WMS.